\documentclass[aps,showpacs,prb]{revtex4}%
\usepackage{amsmath}
\usepackage{amssymb}
\usepackage{graphicx}
\usepackage{amsfonts}%
\newtheorem{theorem}{Theorem}

\newtheorem{lemma}[theorem]{Lemma}

\newenvironment{proof}[1][Proof]{\noindent\textbf{#1.} }{\ \rule{0.5em}{0.5em}}
\begin{document}
\title{Exact Numerical Solution of the BCS Pairing Problem}
\author{Feng Xu }
\affiliation{Department of Modern Physics, University of Science and Technology of China,
Hefei, 230026, P.R.China}
\author{An Min Wang}
\affiliation{Department of Modern Physics, University of Science and Technology of China,
Hefei, 230026, P.R.China}
\author{Xiao-Dong Yang, and Hao You}
\affiliation{Department of Modern Physics, University of Science and Technology of China,
Hefei, 230026, P.R.China}

\begin{abstract}
We propose a new simulation computational method to solve the reduced BCS
Hamiltonian based on spin analogy and submatrix diagonalization. Then we
further apply this method to solve superconducting energy gap and the results
are well consistent with those obtained by Bogoliubov transformation method.
The exponential problem of $2^{N}$-dimension matrix is reduced to the
polynomial problem of $N$-dimension matrix. It is essential to validate this
method on a real quantum computer and is helpful to understand the many-body
quantum theory.

\end{abstract}
\keywords{BCS theory, quantum simulation}
\pacs{74.20.Fg, 03.67.Lx}
\maketitle










BCS theory\cite{bcs} and its subsequent extension is a well established theory
to explain the mechanism of superconducting property. With two gross
simplifications: the free electron approximation and the effective interaction
approximation\cite{solid state}$^{,}$ \cite{taylor}, a simplified BCS model is
obtained and described by the reduced BCS Hamiltonian. There has been much
work on solving this Hamiltonian. The mean field method is exact in the limit
of large number of electrons where fluctuation can be neglected but disabled
in the case of small number of electrons. Since Richardson's
work\cite{richardson} in the 60's to now, the exactly solvable BCS Hamiltonian
attracts much attention in connection with the problems in different areas of
physics such as superconductivity, nuclear physics, physics of ultrasmall
metallic grains.

Recently in \textsl{L.-A. Wu et al.}'s paper\cite{simulation} an NMR
experiment scheme performing a polynomial-time simulation of pairing model was
reported. Based on this work we propose an explicit theory method to
diagonalize the reduced BCS Hamiltonian through the spin analogy and submatrix
diagonalization. Compared with the conventional method it is more useful in
solving practical problem. The problem is solved in the spin space, which is
convenient related to the qubit system. It gives a senseful alive method,
quantum simulation, instead of the numerical diagonalization calculation. And
it shows the potential to solve many-body problem by quantum computer. In fact
more and more people concentrate on the research of simulating other physics
systems by quantum computer\cite{introduce qs}. The experimental quantum
simulations about quantum harmonic oscillator\cite{somaroo}, three-spin
artifical Hamiltonian\cite{Tseng} and migration of excitation in a
one-dimensional chain\cite{Khitrin} \textsl{et al.} have been realized.
Recently a relative experiment is performed to get the eigenvalues of the BCS
Hamiltonian through selecting a proper initial state and realizing Hamiltonian
evolution.\cite{yang}\cite{wang0410007}

The exact solvable model, \textsl{i.e.} the reduced BCS Hamiltonian considered
in this paper is\cite{taylor}$^{,}$\cite{simulation}$^{,}$\cite{li}:%

\begin{equation}
H_{\mathrm{BCS}}=\sum_{m=1}^{N}\frac{(\varepsilon_{m}-\varepsilon_{F})}%
{2}(n_{m}+n_{-m})-V\sum_{m,l=1}^{N}c_{m}^{\dagger}c_{-m}^{\dagger}c_{-l}c_{l}
\label{bcsh}%
\end{equation}
where $n_{\pm m}\equiv c_{\pm m}^{\dagger}c_{\pm m}$ are the electron number
operators, $c_{m}^{\dagger}(c_{m})$ is the fermionic creation (annihilation)
operator. The coupling coefficient is simplified as a constant $V$\cite{solid
state}$^{,}$\cite{couple constant}. Note that the summation indexes
$m=1,2,\cdots,N$ represent all of relevant quantum numbers, and the electron
pairs are labelled by the the quantum number $m$ and $-m$, according to the
Cooper pair situation where the paired electrons have equal energies but
opposite momenta and spins: $m=(\overrightarrow{k},\uparrow)$ and
$-m=(-\overrightarrow{k},\downarrow)$. Introduce the pair creation operator
$b_{m}^{\dagger}=c_{m}^{\dagger}c_{-m}^{\dagger}$ and the pair annihilation
operator $b_{m}=c_{-m}c_{m}$. So one can write the Hamiltonian (\ref{bcsh})
as\cite{solid}:%

\begin{equation}
H_{\mathrm{BCS}}=\sum_{m=1}^{N}\frac{\xi_{m}}{2}(n_{m}+n_{-m})-V\sum
_{m,l=1}^{N}b_{m}^{\dagger}b_{l} \label{pairh}%
\end{equation}
where $\xi_{m}=\varepsilon_{m}-\varepsilon_{F}$ is the free electron kinetic
energy from Fermi surface ($\varepsilon_{F}$ is the Fermi energy). There are
two possible cases for every pair state $m$: \textquotedblleft occupation" and
\textquotedblleft empty", which are denoted respectively by:
\begin{equation}
\chi_{1}=\binom{1}{0},\quad\chi_{0}=\binom{0}{1}%
\end{equation}
where the spin up state $\chi_{1}$ indicates \textquotedblleft occupation" and
the spin down state $\chi_{0}$ indicates \textquotedblleft empty". Obviously,
$\displaystyle\frac{1}{2}(\sigma_{x}-i\sigma_{y})\chi_{1}=\chi_{0}$ and
$\displaystyle\frac{1}{2}(\sigma_{x}-i\sigma_{y})\chi_{0}=0$, then we can get
the so-called spin-analogy corresponding of the pair annihilation operator
$b_{m}$ as
\begin{equation}
b_{m}\Rightarrow%
\begin{pmatrix}
0 & 0\\
1 & 0
\end{pmatrix}
=\frac{1}{2}\left(  \sigma_{x}^{(m)}-i\sigma_{y}^{(m)}\right)  =\sigma_{m}^{-}
\label{bx}%
\end{equation}
In the same way, the spin-analogy corresponding of the pair creation operator
$b_{m}^{\dagger}$ becomes
\begin{equation}
b_{m}^{\dagger}\Rightarrow%
\begin{pmatrix}
0 & 1\\
0 & 0
\end{pmatrix}
=\frac{1}{2}\left(  \sigma_{x}^{(m)}+i\sigma_{y}^{(m)}\right)  =\sigma_{m}^{+}
\label{bk+}%
\end{equation}
From the pair number operator $n_{m}+n_{-m}$ has the eigenvalue 2 (which
represents the electron number in every Cooper pair) when operating on
$\chi_{1}$, and 0 when operating on $\chi_{0}$, it follows that
\begin{equation}
n_{m}+n_{-m}\Rightarrow%
\begin{pmatrix}
2 & 0\\
0 & 0
\end{pmatrix}
=1+\sigma_{z}^{(m)} \label{nk}%
\end{equation}
In fact, the fermionic pair operators satisfy the commutation algebra:
$sl(2)=\left\{  b_{m},b_{m}^{\dagger},n_{m}+n_{-m}-1\right\}  $, \textsl{i.e.}
$sl(2)=\left\{  \sigma_{m}^{-},\sigma_{m}^{+},\sigma_{m}^{z}\right\}  $. From
formulas (\ref{bx})-(\ref{nk}) one can express $H_{\mathrm{BCS}}$ in terms of
the spin operators:%

\begin{align*}
H_{\mathrm{spin}}  &  \Rightarrow\sum_{m=1}^{N}\frac{\xi_{m}}{2}\left(
1+\sigma_{z}^{(m)}\right)  -V\sum_{m,l=1}^{N}\frac{1}{2}\left(  \sigma
_{x}^{(m)}+i\sigma_{y}^{(m)}\right)  \frac{1}{2}\left(  \sigma_{x}%
^{(l)}-i\sigma_{y}^{(l)}\right) \\
&  =\sum_{m=1}^{N}\frac{\epsilon_{m}}{2}\left(  1+\sigma_{z}^{(m)}\right)
-\frac{V}{2}\sum_{m<l=1}^{N}\left(  \sigma_{x}^{(m)}\sigma_{x}^{(l)}%
+\sigma_{y}^{(m)}\sigma_{y}^{(l)}\right)
\end{align*}
where $\epsilon_{m}=\xi_{m}-V$.

In fact the spin analogy of the BCS Hamiltonian is well known and exact
diagonalization of the pairing model in the spin space has been carried out in
several previous works\cite{other dia}. In this paper we propose a
computational simulation method which is potential to realize in future with
the development of quantum computer. The primary advantage of our method lies
in the practical realization in experiments. Especially we can solve
superconducting energy gap by this simulation method conveniently as following
paragraphs. It is more practicably than other solution of energy gap, because
it can simplify a $2^{N}$-dimension problem to an $N$-dimension problem. We
know eigenvalues may not be solvable for high dimension matrix in principle.
Now a $2^{N}$-dimension problem, exponential problem(EP) can be simplified an
$N$-dimension problem, polynomial problem(PP). In the following part we will
describe how to transform EP to PP in detail.

Firstly the total Hamiltonian $H_{\mathrm{spin}}$ will be expressed as the
direct-sum of a set of submatrices\cite{simulation}.
\begin{equation}
H_{\mathrm{spin}}=H_{sub0}\oplus H_{sub1}\oplus H_{sub2}\oplus\cdots\oplus
H_{subN} \label{dirsum1}%
\end{equation}
The system states with the same spin-up state number form an absolute
subspace. The subscripts $sub0,sub1,\cdots subN$ representing the number of
the spin-up state in the corresponding subspace are respectively
$0,1,\cdots,N$. Secondly we will prove that the eigenvalues of $H_{sub1}$ in
sub1 submatrix justly are the eigenvalues of $H_{\mathrm{BCS}}.$

\bigskip Note the $2^{N}\times2^{N}$ operator as%
\begin{equation}
h(N,m)=\mathbf{I}^{\otimes(m-1)}\otimes%
\begin{pmatrix}
1 & 0\\
0 & 0
\end{pmatrix}
\otimes\mathbf{I}^{\otimes(N-m)}\text{ \ }(m=1,2,\cdots,N) \label{h(n,m)}%
\end{equation}
whose $i$-th diagonal element is noted as $h(N,m)[i]$ $(i=1,2,....,N)$. It is
easy to see the non-diagonal elements of $h(N,m)$ are zero.

\bigskip

\begin{lemma}
For the Hamiltonian as $h(N,m)=\mathbf{I}^{\otimes(m-1)}\otimes%
\begin{pmatrix}
1 & 0\\
0 & 0
\end{pmatrix}
\otimes\mathbf{I}^{\otimes(N-m)}$, the $(2^{N}-2^{N-i})$-th element satisfies%
\begin{equation}
h(N,m)[2^{N}-2^{N-i}]=\delta_{im}\text{ \ \ \ \ }(i,m=1,2,\cdots,N)
\label{delta im}%
\end{equation}

\end{lemma}

\begin{proof}
We will prove this lemma by mathematic induction.

\bigskip When $N=3$, it is easy to get three matrices, $m$ value is 1, 2, 3 respectively.%

\[
h(3,1)=%
\begin{pmatrix}
1 &  &  &  &  &  &  & \\
& 1 &  &  &  &  &  & \\
&  & 1 &  &  &  &  & \\
&  &  & ^{\prime}1^{\prime} &  &  &  & \\
&  &  &  & 0 &  &  & \\
&  &  &  &  & ^{\prime}0^{\prime} &  & \\
&  &  &  &  &  & ^{\prime}0^{\prime} & \\
&  &  &  &  &  &  & 0
\end{pmatrix}
,h(3,2)=%
\begin{pmatrix}
1 &  &  &  &  &  &  & \\
& 1 &  &  &  &  &  & \\
&  & 0 &  &  &  &  & \\
&  &  & ^{\prime}0^{\prime} &  &  &  & \\
&  &  &  & 1 &  &  & \\
&  &  &  &  & ^{\prime}1^{\prime} &  & \\
&  &  &  &  &  & ^{\prime}0^{\prime} & \\
&  &  &  &  &  &  & 0
\end{pmatrix}
,h(3,3)=%
\begin{pmatrix}
1 &  &  &  &  &  &  & \\
& 0 &  &  &  &  &  & \\
&  & 1 &  &  &  &  & \\
&  &  & ^{\prime}0^{\prime} &  &  &  & \\
&  &  &  & 1 &  &  & \\
&  &  &  &  & ^{\prime}0^{\prime} &  & \\
&  &  &  &  &  & ^{\prime}1^{\prime} & \\
&  &  &  &  &  &  & 0
\end{pmatrix}
\]
Here the non-diagonal elements are all 0. These diagonal elements with
$^{\prime}$ $^{\prime}$ are $h(3,m)[2^{3}-2^{3-i}]$. It is easy to validate
$h(3,m)[2^{3}-2^{3-i}]=\delta_{im}$ and the last diagonal element
$h(3,m)[2^{3}]$ is zero. If when $N=L$,
\begin{equation}
h(L,m)[2^{L}-2^{L-i}]=\delta_{im}\text{ \ }(i,m=1,2,\cdots,L) \label{l(1)}%
\end{equation}
is right and
\begin{equation}
h(L,m)[2^{L}]=0 \label{l(2)}%
\end{equation}
then we should examine whether $h(L+1,m)[2^{L+1}-2^{L+1-i}]=\delta_{im}$
$(i,m=1,2,\cdots,L+1)$ is right when $N=L+1$.

We will discuss it in two cases: $m\leqslant L$ and $m=L+1$.

(1) $m\leqslant L$:
\begin{align}
h(L+1,m)  &  =\mathbf{I}^{\otimes(m-1)}\otimes%
\begin{pmatrix}
1 & 0\\
0 & 0
\end{pmatrix}
\otimes\mathbf{I}^{\otimes(L+1-m)}\nonumber\\
&  =h(L,m)\otimes\mathbf{I} \label{h(l+1,m)}%
\end{align}
From Eq.(\ref{h(l+1,m)}) it is easy to see%
\begin{equation}
h(L+1,m)[2x]=h(L+1,m)[2x-1]=h(L+1,m)[x] \label{h(2x)}%
\end{equation}
so%
\[
h(L+1,m)[2^{L+1}-2^{L+1-i}]=h(L,m)[2^{L}-2^{L-i}]=\delta_{im}\text{
\ }(i=1,2,\cdots,L)
\]
Next we should also know the value of $h(L+1,m)[2^{L+1}-1]$. From
Eq.(\ref{l(2)}) and Eq.(\ref{h(2x)}) there is $h(L+1,m)[2^{L+1}%
-1]=h(L,m)[2^{L}]=0$. From above discussion for $m\leqslant L(N=L+1)$, the
equality $h(L+1,m)[2^{L+1}-2^{L+1-i}]=\delta_{im}$ is valid.

(2) $m=L+1$:%
\begin{align*}
h(L+1,L+1)  &  =\mathbf{I}^{\otimes L}\otimes%
\begin{pmatrix}
1 & 0\\
0 & 0
\end{pmatrix}
\\
&  =%
\begin{pmatrix}%
\begin{pmatrix}
1 & 0\\
0 & 0
\end{pmatrix}
&  & \\
& \ddots & \\
&  &
\begin{pmatrix}
1 & 0\\
0 & 0
\end{pmatrix}
\end{pmatrix}
\end{align*}
From the above expressions the $\left(  2^{L+1}-1\right)  $-th diagonal
element $h(L+1,L+1)[2^{L+1}-1]=1$, that is to say%
\[
h(L+1,L+1)[2^{L+1}-2^{L+1-i}]=1\text{ }(i=L+1)
\]
And then for $i\neq L+1$,
\[
h(L+1,L+1)[2^{L+1}-2^{L+1-i}]=0
\]
because the even-th diagonal elements are all zero obviously. So for $m=L+1$
there is also $h(L+1,m)[2^{L}-2^{L-i}]=\delta_{im}$.

From the discussion (1) and (2) we have proved that when $N=L+1$,
$h(L+1,m)[2^{L+1}-2^{L+1-i}]=\delta_{im}$ $(i,m=1,2,\cdots,L+1)$ is valid. So
for $\forall N\geqslant3$ ($N$ is the natural number), there is the equality%
\[
h(N,m)[2^{N}-2^{N-i}]=\delta_{im}\text{ \ \ \ \ }(i,m=1,2,\cdots,N)
\]

\end{proof}

\bigskip

After the preparation we will prove that the eigenvalues of $H_{sub1}$ in sub1
subspace justly are the energy spectrum of quasiparticle excitation of
$H_{\mathrm{BCS}}$. Firstly set a diagonal Hamiltonian as $H^{\mathrm{diag}%
}=\frac{1}{2}\sum_{m=1}^{N}E_{m}(\gamma_{m}^{\dagger}\gamma_{m}+\gamma
_{-m}^{\dagger}\gamma_{-m})$, $\gamma_{m}^{\dagger}\gamma_{m}$ and
$\gamma_{-m}^{\dagger}\gamma_{-m}$ are the quasiparticle number operators.
According to the previous analogy rule of number operators Eq.(\ref{nk}) the
spin-analogy form of $H^{\mathrm{diag}}$ is%
\[
H_{\mathrm{spin}}^{\mathrm{diag}}=\sum_{m=1}^{N}E_{m}h(N,m)
\]
$H_{\mathrm{spin}}^{\mathrm{diag}}$'s submatrix in sub1 subspace is denoted as
$H_{sub1}^{\mathrm{diag}}$ and the $i$-th diagonal element of $H_{sub1}%
^{\mathrm{diag}}$ as $H_{sub1}^{\mathrm{diag}}[i]$. We can find the $i$-th
diagonal element of $H_{sub1}^{\mathrm{diag}}$ is the $(2^{N}-2^{N-i})$-th
diagonal element of the total Hamiltonian $H_{\mathrm{spin}}^{\mathrm{diag}}$.
According to Eq.(\ref{delta im})
\begin{align*}
H_{sub1}^{\mathrm{diag}}[i]  &  =H_{\mathrm{spin}}^{\mathrm{diag}}%
[2^{N}-2^{N-i}]\\
&  =\sum_{m=1}^{N}E_{m}h(N,m)[2^{N}-2^{N-i}]\\
&  =\sum_{m=1}^{N}E_{m}\delta_{im}\\
&  =E_{i}%
\end{align*}
So the spin-analogy Hamiltonian of a diagonal BCS Hamiltonian
$H^{\mathrm{diag}}$\ in sub1 subspace has the same eigenvalues as the diagonal
BCS Hamiltonian's energy spectrum. We can deduce further that the eigenvalues
of $H_{sub1}$ are justly the eigenvalues of $H_{\mathrm{BCS}}$ whether which
is diagonal or not, because $H_{\mathrm{BCS}}$ can be written as the diagonal
form like $H^{\mathrm{diag}}$ generally, i.e. $H_{\mathrm{BCS}}\longrightarrow
\frac{1}{2}\sum_{m=1}^{N}E_{m}(\gamma_{m}^{\dagger}\gamma_{m}+\gamma
_{-m}^{\dagger}\gamma_{-m})+\cdots$, while we need not care about how to
obtain the diagonal form Hamiltonian. Consequently it implies that if we have
diagonalized $H_{sub1}$ we can get the energy spectrum of quasiparticle
excitation of $H_{\mathrm{BCS}}$ and get energy gap further avoiding complex
computation. One of the classical methods to diagonalize $H_{\mathrm{BCS}}$,
Bogoliubov transformation method is available under the mean-field
approximation, which is not exact especially in the case of limited $N$.
Another famous method about the exact solution of BCS Hamiltonian has been
proposed in the 60's by Richardson\cite{richardson}. He considered the system
with $M\leqslant N$ pairs electrons and constructed a set of operators
$\mathbf{R}_{j}$ $(j=1,\cdots,N)$ commuted with $H_{\mathrm{BCS}}$, finally
gave the expression of the eigenvalues $\lambda_{j}$ of $\mathbf{R}_{j}$
through solving the $M$\ coupled algebraic equations. $\lambda_{j}$ is not yet
the eigenvalue of $H_{\mathrm{BCS}}$. We can more directly and simply give the
energy spectrum of $H_{\mathrm{BCS}}$.

Thus it can be seen the idea of diagonalizing $H_{sub1}$ instead of solving
the eigenvalues of $H_{\mathrm{BCS}}$ directly is better than those classical
ones. Now a key problem that how to get the submatrix $H_{sub1}$ is placed to
the front. The correlative work\cite{wang} in our group has proved that the
general form of $H_{sub1}$:%
\begin{equation}
H_{sub1}[i,i]=\epsilon_{i},\text{ \ \ \ }H_{sub1}[i,j]=-V\text{ \ }%
(i,j=1,2,\cdots,N;\text{ }i\neq j) \label{wang}%
\end{equation}
$H_{sub1}[i,j]$ is the matrix element of $H_{sub1}$.

\bigskip

Finally it is necessary to check the methods in numerical computation. Here we
will compare our solution with the result of the mean-field approximation by
the value of superconducting energy gap $\Delta$ $(T=0K)$. According to the
physics meaning of $\Delta$, the energy required to excite at least a
quasiparticle from the Fermi surface, the energy of the element excitation is
written as $E_{m}=(\xi_{m}^{2}+\Delta^{2})^{1/2}$.\cite{taylor} In fact the
element excitation energy is also the eigenvalue of $H_{\mathrm{BCS}}$. After
getting the eigenvalue of $H_{sub1}$, \textsl{i.e.} the eigenvalue of
$H_{\mathrm{BCS}}$, we can get the value of $\Delta$ by solving equation
$E_{m}=(\xi_{m}^{2}+\Delta^{2})^{1/2}$. But in order to get rid of the effect
of energy zero the equation%
\begin{equation}
(\xi_{1}^{2}+\Delta^{2})^{1/2}-(\xi_{2}^{2}+\Delta^{2})^{1/2}=E_{1}-E_{2}
\label{gapeqnew}%
\end{equation}
is used to solve $\Delta$ in practice, because the difference of eigenvalues
dosen't depend on the energy zero and that we find the energy difference
between the ground and the first excited state $E_{2}-E_{1}$ is by far larger
than $E_{3}-E_{2},E_{4}-E_{3},\cdots,E_{N}-E_{N-1}$ in the course of the
numerical computation. Here $\xi_{1}<\xi_{2}<\cdots<\xi_{N}$ and $E_{1}%
<E_{2}<\cdots<E_{N}$\bigskip. The another kind of solution used in comparing
is the following energy gap equation\cite{li}:%
\begin{equation}
1=\frac{1}{2}V\sum_{m}\frac{1}{\sqrt{\xi_{m}^{2}+\Delta^{2}}} \label{gapeqold}%
\end{equation}

Here we consider the reduced BCS model whose energies are given for simplicity
by $\xi_{m}=m\delta$\cite{83172}$^{,}$\cite{783749}$^{,}$\cite{804542}, here
$\delta$ is the average level spacing which is inversely proportional to the
size of the grains. In the strong coupling regime, corresponding to large
grains or strong coupling constants, $\delta\ll\Delta$. In the weak coupling
region, corresponding to small grains or small coupling constants, $\delta
\gg\Delta$ \cite{83172}. From much research about ultrasmall superconducting
grains\cite{reports}, the mean-field theory is not suitable in the weak
coupling region. It has been proved that the corrections to the mean-field
results are small in large grains become important in the opposite
limit\cite{783749}. So in this paper we carry out the numerical computation in
the first case, $\delta\ll\Delta$. We also take the coupling constants
$V=\lambda\delta$ in order to discuss conveniently. In order to give the
numerical pictures, we suppose the value of $V$ by the rough estimate. From
BCS theory, the Cooper pair lies in the attraction area, \textsl{i.e.}
$0\leq\xi_{m}\leq\hbar\varpi_{D}$\cite{li}. For metal Debye energy
$\hbar\varpi_{D}\sim10^{-2}$eV, we can set $V\thicksim2\times10^{-6}$eV by
rough estimate. The estimate process is put to the later appendix. Another two
variables $\lambda$ and $N$ are taken as the independent variables of the
energy gap.

\begin{figure}[h]
\includegraphics[width=0.5\textwidth]{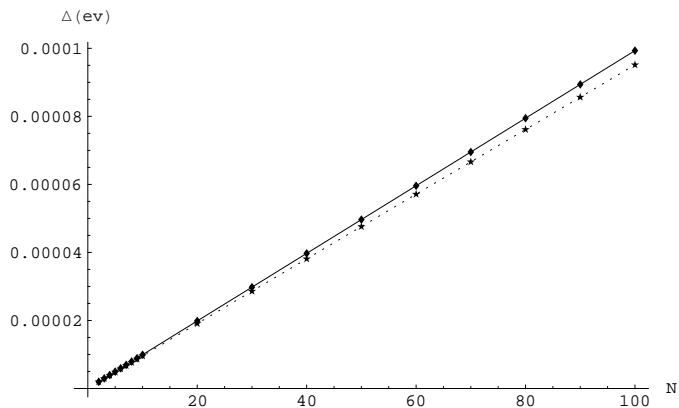}\caption{The energy gap $\Delta$
as a function of the number of energy level, $N$ . We give the comparison
between two results by the different methods. The solid line is the solution
of the energy-gap equation; the dashed line is our result by spin analogy and
diagonalizing submatrix. Here we choose $V\thicksim2\times10^{-6}$eV,
$\lambda=10$ .}%
\label{gap-m}%
\end{figure}

We list our results in diagrams. In FIG.\ref{gap-m} setting $\lambda=10$, the
energy gap is plotted as the function of energy level number, which is the
mono-increasing function of the energy level number $N$.

\begin{figure}[hh]
\includegraphics[width=0.5\textwidth]{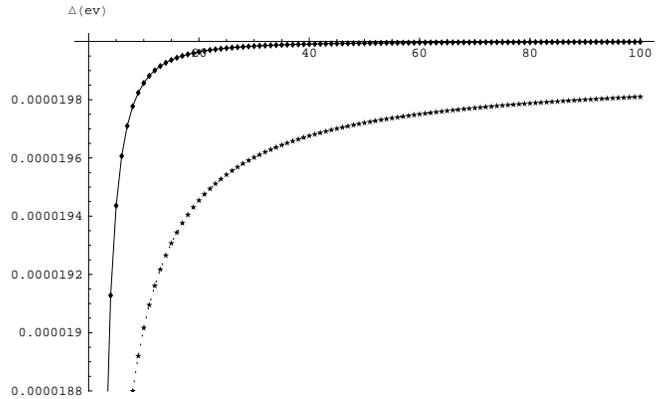}\caption{When $N=20$, Energy
gap $\Delta$ is plotted as the function of $\lambda$. We give the comparison
between two results: the solid line is the solution of the energy-gap
equation; the dashed line is our result by spin analogy and diagonalizing
submatrix. Here we choose $V\thicksim2\times10^{-6}$eV.}%
\label{gap-namda}%
\end{figure}

\noindent The relative error between our result and that of the
Eq.(\ref{gapeqold}) is not more than 5\% in the range from $N=2$ to $N=100$
and fixed $\lambda(=10)$. It shows that the result from our method is well
consistent with the solution of the energy gap equation. In order to check the
universality of this new method, we also give the dependence relation between
the energy gap and the level spacing, see FIG.\ref{gap-namda}.

\noindent It is clear to see $\Delta$ changes gently with $\lambda$ when
$\lambda$ is large enough. That is to say, $\Delta$ is almost independent of
$\delta$ when $\delta$ is small enough. It shows the rationality of $\delta
\ll\Delta$ on the inverse hand. Obviously, in FIG.\ref{gap-namda}, when
$\lambda$ is larger than 80, the relative error is less than 1.1\%.

\begin{figure}[h]
\includegraphics[width=0.5\textwidth]{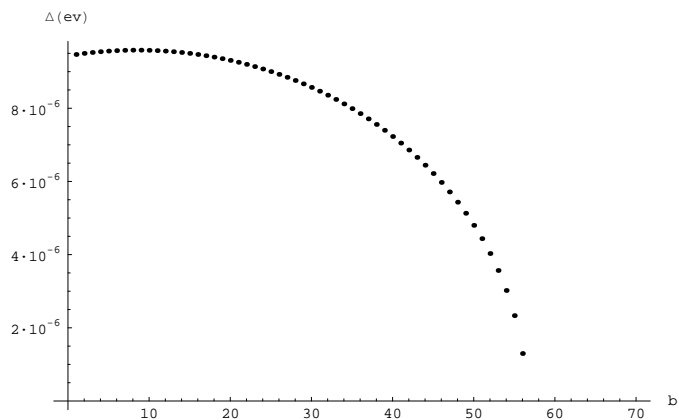}\caption{ When $V\thicksim
2\times10^{-6}$eV, $\lambda=10$, $N=10$ the energy gap $\Delta$ is ploted as
the function of $b$ through the quantum simulation. When $b>56$, there is not
the real root for $\Delta$. The result is obtained by the quantum simulation.
Similar result will be obtained by the energy gap equation Eq.(\ref{gapeqold}%
).}%
\label{gap-b}%
\end{figure}

We also consider a small departure from the fermi surface, that is $\xi
_{m}=\xi_{0}+m\delta$. $\xi_{0}$ is a small value, $\xi_{0}\sim$ $\delta$. In
the following discussion we note $\xi_{0}=b\delta$, $b$ is the natural number.
The small departure from the fermi surface reduces the energy gap and energy
gap is not a real root when the departure reaches a critical value, see
FIG.\ref{gap-b}.

According to above comparison we know two results are consistent well, while
our method to solve the energy spectrum doesn't include approximation, which
indicates that our result includes that obtained by mean field theory and is
superior to it.

In summary, we have proposed an exact numerical simulation method to calculate
the energy spectrum of the reduced BCS Hamiltonian by spin analogy and
diagonalizing submatrix. A numerical computation to verify the validity of our
computational method is given. We make a comparison between our method and
energy gap equation Eq.(\ref{gapeqold}), and two results are well consistent
in numerical computation. By examining the change of the energy gap value
under the change of the parameter, we include the excellent consistency
between the two results by the different methods is independent on the
particular parameter. It implies that one can implement this quantum
simulation on a quantum computer and the result will be believable. Currently
a new experiment about 2-qubit simulation of the pairing Hamiltonian on an NMR
quantum computer has been realized and get the energy spectrum of the pairing
Hamiltonian successfully\cite{yang}. With the development of quantum computer,
especially the manipulation and control of multi-qubit system, this new
simulation computation method has the\ great potential in practical application.\ 

We are grateful Xiaosan Ma, WanQing Niu, Zhao Ningbo, Zhu Rengui and Su
Xiao-Qiang for helpful discussion. This work was founded by the National
Fundamental Research Program of China with No. 2001CB309310, partially
supported by the National Natural Science Foundation of China under Grant No.
60173047 and the Natural Science Foundation of Anhui Province.

\appendix

\section{Appendix: Estimate about $V$}

For metal element $g(0)V\approx0.2\sim0.3$\cite{li}, $g(0)$ is the state
density which have some spin directions on the Fermi surface.
\begin{equation}
\int_{0}^{\infty}2g(\varepsilon)d^{3}\varepsilon=1 \label{renormal}%
\end{equation}
According to the assumption%
\begin{align*}
g(\varepsilon)  &  =0\text{, \ \ \ \ \ if }\varepsilon>\hbar\varpi_{D}\\
g(\varepsilon)  &  =g(0)\text{, \ \ if }\varepsilon\leq\hbar\varpi_{D}%
\end{align*}
formalism(\ref{renormal}) can be written as%
\[
\int_{0}^{\hbar\varpi_{D}}2g(0)d^{3}\varepsilon=1
\]
so from BCS theory, the Cooper pair lies in the attraction area, ie $0\leq
\xi_{k}\leq\hbar\varpi_{D}$. On substitution of $\hbar\varpi_{D}%
\thicksim10^{-2}$eV \cite{li} we can estimate $g(0)\approx10^{5}$, so
$V\thicksim2\times10^{-6}$eV.


\end{document}